\documentclass[twocolumn,showpacs,preprintnumbers,amsmath,amssymb]{revtex4-1}

\usepackage{graphicx}
\usepackage{dcolumn}
\usepackage{bm}

\usepackage{epstopdf}
\begin{document}

\title{Zero-Bias Anomalies in Narrow Tunnel Junctions in the Quantum Hall Regime}

\author{P. Jiang$^{1,2}$, C.-C. Chien$^{1}$, I. Yang$^{1}$, W. Kang$^{1}$, K. W. Baldwin$^{3}$, L. N. Pfeiffer$^{3}$, and  K. W. West$^{3}$}

\affiliation{$^{1}$James Franck Institute and Department of Physics, University of Chicago, Chicago, Illinois 60637\\
$^{2}$Department of Physics, National Taiwan Normal University, Taipei 116, Taiwan\\
$^{3}$Department of Electrical Engineering, Princeton Univeristy, Princeton, NJ 08544}

\begin{abstract}
We report on the study of cleaved-edge-overgrown line junctions with a serendipitously created  narrow opening in an otherwise thin, precise line barrier. Two sets of zero-bias anomalies are observed with an enhanced conductance for filling factors $\nu > 1$ and a strongly suppressed conductance for $\nu < 1$. A transition between the two behaviors is found near $\nu \approx 1$. The zero-bias anomaly (ZBA)  line shapes find explanation in Luttinger liquid models of tunneling between quantum Hall edge states. The ZBA for $\nu < 1$ occurs from strong backscattering induced by suppression of quasiparticle tunneling between 
the edge channels for the $n = 0$ Landau levels.
The ZBA for $\nu > 1$ arises from weak tunneling of quasiparticles between the $n = 1$ edge channels.
\end{abstract}

\maketitle

Studies of tunneling properties of  the edge state of fractional quantum Hall effect (FQHE) states have been instrumental in revealing the unique correlation physics of chiral Luttinger liquids \cite{KaneFisher,Chang03,Wen90a,Wen91,Fendley95}. Recent development in topological quantum computing has rekindled interest in the physics of quasiparticle tunneling in the FQHE \cite{Nayak08}. Tunneling of quasiparticles in  multiply connected interferometers is necessary to create entangled states of quasiparticle wave functions. A better understanding of the edge-state physics is essential to the study of quantum information in the quantum Hall regime.

The chiral Luttinger liquid model of the FQHE  edge state predicts a power-law dependence of the tunneling conductance on the bias voltage, $G \sim V^{\alpha}$, where the exponent $\alpha$ is determined by the bulk FQHE state \cite{KaneFisher,Chang03,Wen90a,Wen91,Fendley95}.  Experimental studies of narrow-constriction tunnel junctions in the FQHE  regime have provided a measure of support of the model. At filling factor $\nu = 1/3$, the tunneling conductance shows the predicted scaling behavior over some ranges of experimental parameters \cite{Milliken96,Roddaro03,Roddaro04}. More recently a study of the narrow-junction tunneling conductance in the second Landau level was able to establish the limiting values of the fractional charge and the interaction parameter for the $\nu = 5/2$ FQHE  state from comparison to the theoretical scaling forms \cite{Radu08}. 

In a tunneling experiment between a $\nu = 1/3$ FQHE edge state and a three-dimensional electronic system, a power-law behavior over an extended range of bias voltages and temperatures was found in support of the chiral Luttinger liquid model \cite{Chang96}. However, later studies revealed that the extended power-low behaviors can be observed at both compressible and incompressible states with the exponent evolving continuously with the bulk filling factor \cite{Grayson98,Hilke01}. These findings have raised a number of unanswered questions regarding  the nature of electron-electron interactions in the edge states of compressible and incompressible states in the FQHE  regime \cite{Chang03,Shytov98,Levitov01}, since the chiral Luttinger liquid model is applicable only to the edge of FQHE states with odd-denominator filling factors. 
In a related experiment, novel, one-dimensional metallic and insulating states have been detected at bent quantum Hall junctions\cite{Grayson07,Steinke08}.

In this paper, we report on study of quantum Hall line junctions, each with some defect in an otherwise perfect line barrier.  In addition to the inter-edge interaction present along a thin tunnel barrier, two counterpropagating edge states are joined at an opening in the barrier that serves as a narrow constriction. For bulk filling factors $\nu > 1$, a zero-bias anomaly (ZBA) with a sharp enhancement of the conductance is  detected. In the lowest Landau level, the ZBA exhibits a strong suppression of the conductance. A clear delineation of the two behaviors points to a quantum phase transition in the tunneling characteristics near $\nu \approx 1$. These ZBAs clearly demonstrate the highly correlated properties of edge state in the quantum Hall regime. While the $\nu < 1$ regime is dominated by strong backscattering, the $\nu > 1$ regime is described by weak tunneling of quasiparticles between 
the edge channels for the $n = 1$ Landau levels.

The junctions were fabricated using cleaved-edge-overgrowth technique to create a sharp rectangular barrier on the plane of two-dimensional electron systems \cite{Pfeiffer90,Kang00}. Two side-by-side, nearly identical 10 $\mu$m-wide strips of two-dimensional electrons are separated from each other by  a 8.8-nm-thick Al$_{0.3}$Ga$_{0.7}$As barrier. In this experiment we studied several samples with a bulk electron density $n_\text{b}  \approx 1.85\times 10^{11}$ cm$^{-2}$. The (110) monitor wafers yielded a typical mobility of $\sim 5\times 10^{5}$ cm$^{2}$/Vs. Results reported in this paper were obtained from a set of samples with anomalous large conductance compared to that of typical junctions. 
The bulk filling factors were determined from independent measurements
of the Hall resistance of the 2DES  part of the junction.
 From reproducible and consistent data 
 we hypothesize that  some imperfection or impurity in the barrier has created a narrow opening through which the predominant electron transport occurs.

\begin{figure}
\includegraphics[width=3.25in]{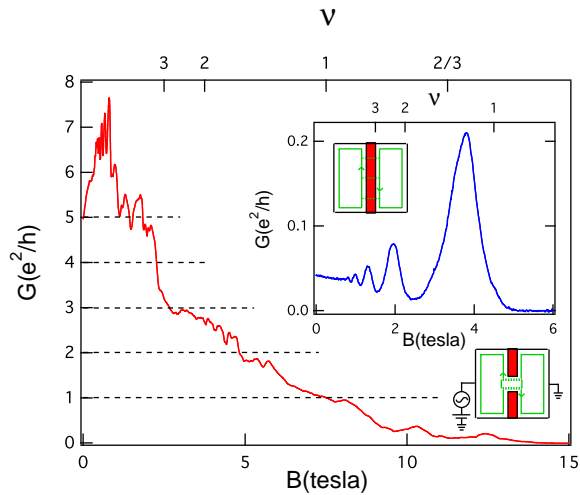}
\caption{\label{fig:fig1}
Zero-bias conductance of a line junction with an imperfect barrier at a temperature of 0.3 K. The bulk electron density $n_\text{b} \approx 1.85\times 10^{11}$ cm$^{-2}$. The bulk filling factors $\nu$ are illustrated at the top of the figure. Inset: Zero-bias conductance of a line junction with a wholesome barrier.}
\end{figure}

In our experiment we measured the differential conductance $G$ across the tunnel junction. 
Fig.~\ref{fig:fig1} illustrates
$G$ at zero-bias voltage as a function of the perpendicular magnetic field  $B$ for a quantum Hall line junction with an imperfect tunnel barrier at a temperature of 0.3 K.  $G$ is approximately $5e^{2}/h$ under $B = 0$. Between $B = 0$ and 1  T,  $G$ increases slightly to $\sim 7e^{2}/h$ and exhibits some weak oscillations. Above 1  T, $G$ decreases quickly as $B$ is increased to 15  T . Some fluctuations and weak plateaus are detected at intermediate $B$, and $G$ becomes smaller than $e^2/h$ beyond 7.5  T  ($\nu \approx 1$). As a comparison, the zero-bias $G$ of a line junction with a wholesome barrier is shown in the inset. $G$ is considerably smaller in a wholesome junction ($G \sim 0.05 e^2/h$ at $B = 0$), and exhibits a $1/B$-periodic behavior with a maximum of $0.2 e^2/h$ near $\nu = 1.3$. For $\nu < 1$, tunneling is completely suppressed and $G$ becomes vanishingly small ( $< 0.01e^{2}/h$). $G$ maxima for the wholesome junctions are interpreted in terms of energy level crossings of the edge-state electrons \cite{Kang00,Ho94,Takagaki00,Mitra01,Nonoyama02}, a sequence of quantum phase transitions \cite{Yang04}, and the transport through a defect in the barrier \cite{Kim03}.

\begin{figure}
\includegraphics[width=2.5in]{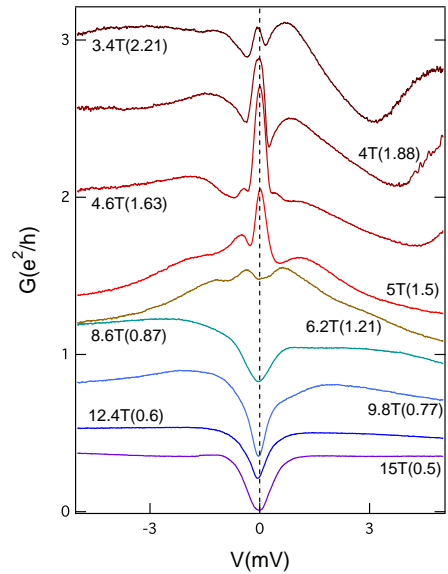}
\caption{\label{fig:fig2}
Bias-voltage dependence of the conductance of a line junction with an imperfect barrier under different magnetic fields. The temperature is 0.3 K.  The numbers in parenthesis indicate the corresponding bulk filling factors.}
\end{figure}

Since the quantum Hall line junctions were originally designed to be in the weak tunneling regime,  the large $G$ in these samples indicates presence of a defect in the barrier.
Defect tunneling was also observed in another experiment with a cleaved-edge-overgrown barrier between a bulk doped electrode and a FQHE edge \cite{Grayson01}. In our experiment,
if multiple defects were present, there should be an Aharonov-Bohm effect due to a possibility of enclosed orbits between the defect sites. Absence of Aharonov-Bohm oscillations implicates that the transport in these samples is dominated by a single, highly tunneling site in the barrier. 
Small fluctuations in  $G$ near $B = 0$ appears to be associated with Shubnikov-de Haas oscillations of the bulk of the sample.

Fig.~\ref{fig:fig2} illustrates the evolution of $G$ vs $V$, the bias voltage, in the integer and fractional quantum Hall regimes. For $\nu > 1$ the most prominent feature is the dramatic enhancement of the zero-bias $G$. The zero-bias $G$ peaks are found between $\nu \approx 2.3$ and 1.3 with the strongest occurring around $\nu = 1.63$, where $G$ at zero bias exceeds $0.7 e^2/h$ relative to the $G$ background measured at larger $|V|$, and the half-width of the peak is around 0.35 mV. Under larger $|V|$, $G$ is weakly dependent on $V$ and decreases gradually with increasing $B$. Around $\nu = 1.2$, the zero-bias $G$ evolves from a peak to a valley. For $\nu < 1$ the zero-bias $G$ is greatly suppressed and becomes close to zero at $\nu = 1/2$ ($B \approx 15$ T). $G$ is nearly independent on $V$ for $|V|$ above $\sim$ 1 mV. 
For $\nu < 1$, $G$ at large $|V|$ approaches the Hall conductance ($G_{xy}$). 
Such an ohmic behavior at large $|V|$ suggests  that the observed nonlinearities at small $|V|$ are a sign of the low-energy correlation of the edge states. 
For $\nu > 1$, there is additional nonlinearity present at large $|V|$ responsible for the non-monotonic background and the conductance larger than $G_{xy}$. The origin of this non-monotonicity for higher Landau levels is unclear but likely involves the modified single-particle energy spectrum near the barrier \cite{Ho94,Takagaki00,Mitra01}.

\begin{figure}
\includegraphics[width=3.25in]{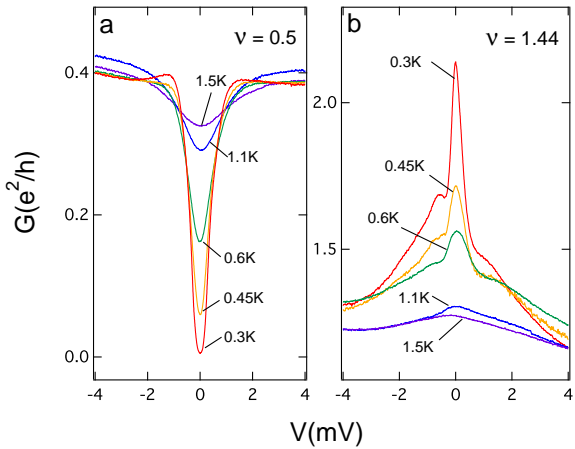}
\caption{\label{fig:fig3}
Temperature dependence of (a) the zero-bias conductance valley at $\nu \approx 0.5$ ($B$ = 15  T) and (b)  the zero-bias conductance peak at $\nu \approx 1.44$ ($B$ = 5.2  T).}
\end{figure}

Representative temperature dependences of the ZBAs in the fractional and integer quantum Hall regimes are illustrated respectively in Figs.~\ref{fig:fig3}a and  \ref{fig:fig3}b. At $\nu \approx 1/2$ the zero-bias $G$ reaches less than $0.005e^{2}/h$  with a half-width of the minimum of around 0.79 mV.  As the temperature is increased, the $G$ minimum is gradually lifted and approaches the large-bias background above 1.5 K.  As an example in the integer regime with $1 < \nu < 2$, on the other hand, the zero-bias $G$ at $\nu \approx 1.44$ at 0.3 K exhibits a prominent $\sim70\%$ enhancement relative to the high-temperature data with the zero-bias peak disappearing above 1 K.

\begin{figure}
\includegraphics[width=3in]{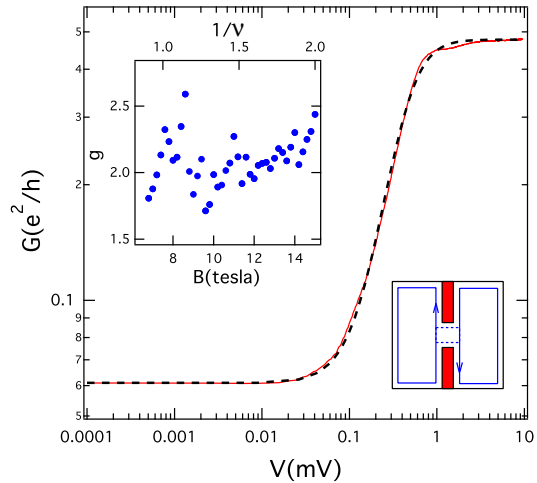}
\caption{\label{fig:fig4}
Log-log plot of conductance vs bias voltage (solid line) in the lowest Landau level at $\nu  \approx0.53$ ($B=14.2$ T) measured at 0.3 K, along with the fit (dashed line) of the data based on the theory of Chamon-Fradkin with $g = 2.06$ and Kondo temperature $T_\text{K} =4.9$ K. Upper inset: Luttinger parameter $g$ as a function of the magnetic field. Lower inset: edge channel for $\nu < 1$.
}
\end{figure}

The small $G$ at zero bias in the lowest Landau level is consistent with strong backscattering induced by suppression of quasiparticle tunneling due to the narrowness of the constriction  (lower inset, Fig.~\ref{fig:fig4}). This is the likely difference between our experiment and the work of Roddaro \textit{et al.} with wider junctions \cite{Roddaro03,Roddaro04}.
We analyze our data for $\nu < 1$ in terms of the Luttinger liquid model of tunneling into the edge of a FQHE state in the lowest Landau level through an impurity proposed by Chamon and Fradkin \cite{Chamon97}. In their model, tunneling from an electron gas to a FQHE edge is mapped into tunneling between two chiral Luttinger liquids with the Luttinger parameter, $g$. In the exactly solvable form of the conductance, a power-law behavior is followed by an ohmic region above a characteristic energy defined as the Kondo temperature, $T_\text{K}$. The mapping in the model works in such a way that the backscattering in the Luttinger theory corresponds to the transmission across the barrier in the experiment.

The theoretical conductance of the Chamon-Fradkin model \cite{Chamon97} provides fair to excellent fits of the experimental data, depending on the filling factor. Fig.~\ref{fig:fig4}
shows a representative plot of $G$ vs $V$ at $B = 14.2$  T  ($\nu \approx 0.53$) and the fit based on the model. A power-law regime at intermediate $V$ connects two ohmic regions at low and high $V$. The crossover behavior  is well approximated by a theoretical curve with $g =  2.06$. At high $V$, $G$ saturates around $0.5e^{2}/h$ with the saturation voltage set by $T_\text{K} = 4.9$ K. The inset shows $g$ for the conductance data for $\nu$ between 0.5 and 1.2.  $g$ is weakly dependent on $B$ with $g \approx 2$. This relative insensitivity of $g$ is consistent with the significance of Coulomb interaction in the determination of the Luttinger parameter in the lowest Landau level
 \cite{Chamon97, Chang03, Shytov98, Levitov01}. 
 A large scatter in $g$ is found near $\nu  = 1$, which reflects the limited dynamic range for fitting due to increasing asymmetry in the ZBAs around $\nu = 1$.

\begin{figure}
\includegraphics[width=2.25in]{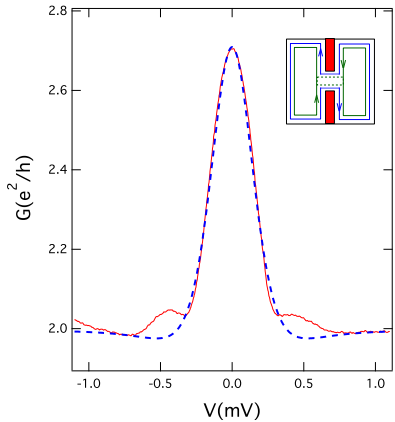}
\caption{\label{fig:fig5}
Fit (dashed curve) of the zero-bias conductance anomaly at $\nu \approx 1.63$ ($B = 4.6$ T) (solid curve) based on the theory of X. G. Wen with interaction parameter $g = 0.27$ and charge $e^{*} = 0.12e$. Inset:  edge channels for $\nu > 1$ where primary tunneling occur between the $n = 1$ edge channels.
}
\end{figure}

The zero-bias $G$ peaks found for $\nu > 1$ may be explained by  the theory of weakly tunneling quasiparticles proposed by Wen \cite{Wen91a}. In comparison to the lowest Landau level, the interedge tunneling in higher Landau levels occurs with a perfect transmission of the $n = 0$ edge channels (inset, Fig.~\ref{fig:fig5}). The enhanced  zero-bias $G$  occurs from weak tunneling of quasiparticles between the $n = 1$ edge channels through the opening in the barrier. Fig.~\ref{fig:fig5} shows the fit of $G$ at $B = 4.6$  T  ($\nu   \approx 1.63$) using Wen's model. 
We note that the asymmetric background complicates obtaining a good value of $\chi^{2}$. However, the ZBA can be well reproduced by concentrating on the region of $\pm 1.5$ mV around zero bias. A good fit of the $G$ peak was obtained with the interaction parameter $g = 0.27$ and the effective charge $e^{*} = 0.12e$. The knees in $G$ near $\pm 0.5$ mV may arise from the fact that the transport for $\nu > 1$ involves multiple edge channels. Landau level mixing and reversed spins likely play a substantial role beyond the lowest Landau level.

The striking contrast between the ZBAs for  $\nu > 1$ and $\nu < 1$ distinguishes the physics of inter-edge correlation in the integer and fractional regimes. The ZBAs, along with the theoretical descriptions of Chamon-Fradkin and Wen models, demonstrate the Luttinger liquid properties of the quantum Hall edge states. With an increasing magnetic field, 
a quantum phase transition occurs around $\nu = 1$ into the regime where the $G$ peak at zero bias becomes a minimum. 

A notable feature of the data is the apparent lack of distinction in the conductance line shape between the compressible and incompressible bulk quantum Hall states for both $\nu > 1$ and $\nu < 1$. For $\nu < 1$, the similar behavior was also found in the experiment of tunneling between a Fermi liquid and the edge state in the lowest Landau level, where a continuous sequence of exponents as a function of the bulk filling factor was observed \cite{Chang03,Grayson98,Hilke01}. This may be a generic feature associated with the hard confining potentials produced in cleaved-edge overgrown junctions. Our experiment circumstantially supports the case of a hard confining potential playing a key role in apparent decoupling of the bulk state from the edge state in cleaved-edge-overgrown junctions.

Although the Chamon-Fradkin and Wen models respectively provide explanations for the ZBAs in $\nu < 1$ and $\nu > 1$ in terms of quasiparticle tunneling, the apparent decoupling between the bulk and the edge physics raises a question on whether the observed nonlinearities can be interpreted in terms of quasiparticle transport. To this end, a greater understanding of the edge-state properties, especially those connected to possible edge-state reconstruction in  the cleaved-edge-overgrown structures \cite{Wan03,Yang03}, is necessary.

In summary,  we have studied a pair of parallel edge states in quantum Hall line junctions coupled through a defect in the barrier. 
Analysis of zero-bias anomalies in terms of the Chamon-Fradkin model shows that the $\nu < 1$ region is dominated by strong backscattering.  On the other hand, the $\nu > 1$ region is phenomenologically described by weak tunneling of quasiparticles between the $n = 1$ edge channels. With an increasing magnetic field,  
a quantum phase transition occurs around $\nu = 1$ into the regime where the conductance peak at zero bias is strongly suppressed. 

We thank N. Cooper, E. Fradkin, E. Kim, A. Ludwig, A. MacDonald, E. Papa, and X. G. Wen for useful discussions. The work at the University of Chicago is supported by NSF MRSEC and Microsoft Corporation Project Q.

\end{document}